\documentclass[amssymb,preprint,onecolumn,floats,amsmath,superscriptaddress,linenumbers]{revtex4}

\usepackage[T1]{fontenc}       
\usepackage{color}             
\usepackage{bm}
\usepackage{graphicx}
\usepackage{amssymb}
\usepackage{amsfonts}
\usepackage{amsmath}
\usepackage{epstopdf}
\usepackage{multirow}
\usepackage{color}
\usepackage{tikz}
\usetikzlibrary{fit, matrix}

\def \FUW{Institute of Experimental Physics, Faculty of Physics, University
of Warsaw, Pasteura St. 5, 02-093 Warsaw, Poland}
\def \CNBCH{Biological and Chemical Research Centre, University of Warsaw,
\.{Z}wirki i Wigury St. 101, 02-089 Warsaw, Poland}

\begin{document}

\title{Long-Distance Coupling and Energy Transfer between Exciton States in Magnetically Controlled Microcavities}

\author{Maciej~\'{S}ciesiek}\affiliation{\FUW}
\author{Krzysztof~Sawicki}\affiliation{\FUW}
\author{Wojciech~Pacuski}\affiliation{\FUW}
\author{Kamil~Sobczak}\affiliation{\CNBCH}
\author{Tomasz~Kazimierczuk}\affiliation{\FUW}
\author{Andrzej~Golnik}\affiliation{\FUW}
\author{Jan~Suf\mbox{}fczy\'{n}ski}\email{Jan.Suffczynski@fuw.edu.pl}\affiliation{\FUW}
\date{\today}

\begin{abstract}
Coupling of quantum emitters in a semiconductor relies, generally, on
short-range dipole-dipole or electronic exchange type interactions.
Consistently, energy transfer between exciton states, that is, electron-hole
pairs bound by Coulomb interaction, is limited to distances of the
order of 10~nm. Here, we demonstrate polariton-mediated coupling
and energy transfer between excitonic states over a distance exceeding
2~$\mu$m. We accomplish this by coupling quantum well-confined excitons
through the delocalized mode of two coupled optical microcavities. Use
of magnetically doped quantum wells enables us to tune the confined
exciton energy by the magnetic field and in this way to control the
spatial direction of the transfer. Such controlled, long-distance
interaction between coherently coupled quantum emitters opens
possibilities of a scalable implementation of quantum networks and
quantum simulators based on solid-state, multi-cavity systems.\newline
\end{abstract}

\maketitle

\section*{Introduction}

Energy transfer between quantum emitters, such as excitons in a semiconductor, relies on a coupling through short-range dipole-dipole (F\"{o}rster)\cite{Forster:1948} or electronic exchange (Dexter)\cite{Dexter:JChemPhys1953} type interactions. The distance of these interactions is typically of the order of 10~nm in III-V or II-VI semiconductors.\cite{Heimbrodt:PRB1998,Itskos:PRB2007} As shown in the pioneering work by Agranovich~{\it
et~al.},\cite{Agranovich:SSC1997} the spatial range of energy transfer in
semiconductor systems can be enhanced utilizing the light-matter coupling
effects. In the strong light-matter coupling regime the exciton and the optical mode of the microcavity exchange energy in a reversible way, which leads to a superposition of their wave functions and the emergence of a new eigenstate called an exciton-polariton \cite{Weisbuch:PRL1992,Dang:PRL1998,
Kasprzak:Nature2006,Amo:NatPhys2009, Kavokin:2017, Klembt:Nature2018,Stepanov:NatComm2019}. When several excitonic states are strongly coupled to a common optical mode, their wave functions hybridize enabling their mutual coupling. In this way polariton-mediated transfer of energy between distant excitons is possible. The energy transfer remains efficient as long as the strong coupling conditions and hybridization of the initial and final excitonic states of the process are maintained.\cite{Zhong:Angewandte2017} As such, it enables an enhancement of the energy transfer range by orders of magnitude with respect to the F\"{o}rster limit.\cite{Coles:NatureMat2014,Feist:PRL2015,Zhong:AngewandteChem2016, Zhong:Angewandte2017,Jayaprakash:LSA2019}

\begin{figure}
   \includegraphics[width=0.53\linewidth]{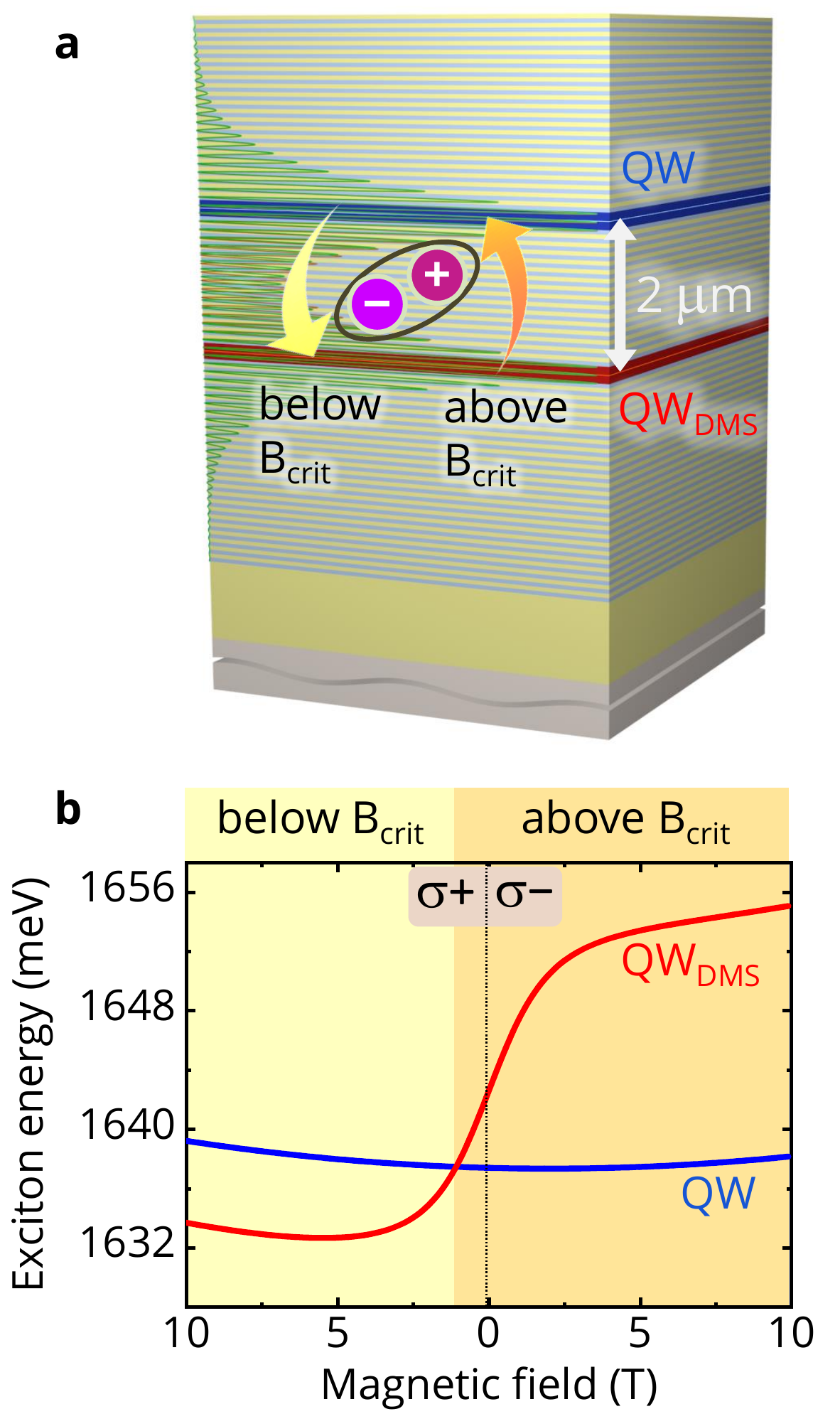}
\caption{\textbf{Schematic view of magnetic field controlled, polariton-mediated energy transfer between exciton states over
2~$\mu$m.} {\bf a} Coupled microcavity structure hosting spatially delocalized
symmetric ($C_{\text{S}}$) and antisymmetric ($C_{\text{AS}}$) optical modes
(respective distributions of the squared electric field shown). Three
non-magnetic quantum wells ($QW$) are placed in the upper and three magnetically
doped quantum wells ($QW_{\text{DMS}}$) are placed in the lower microcavity.
Under strong coupling conditions a four-level polariton system emerges
inducing hybridization of excitons in the $QW_{\text{DMS}}$ and $QW$.
Long-distance transfer of energy between the $QW$ and $QW_{\text{DMS}}$ is
possible thanks to the optical mode mediated coupling of the initial and
final excitonic states of the process. {\bf b} Energy levels of the $QW$ {\color{black}(blue line)} and
$QW_{\text{DMS}}$ {\color{black}(red line)} as a function of the magnetic field. The energy of the excitons in
the $QW_{\text{DMS}}$ can be tuned below or above the energy of the
excitons in the $QW$; the crossing occurs at $B_{crit}$. Below $B_{crit}$, exciton density transfer assisted by energy relaxation occurs predominantly from the $QW$ to the $QW_{\text{DMS}}$, while above $B_{crit}$ the transfer direction is reversed.}
  \label{theidea}
\end{figure}

In this work, we report on magnetic field controlled, polariton-mediated energy transfer between 2D exciton states over a distance as large as 2.15~$\mu$m. We achieve this in structures containing (Cd,Zn)Te and (Cd,Mn,Zn)Te quantum wells strongly coupled to the modes of two coupled microcavities separated by a semi-transparent Bragg mirror (see Figure~\ref{theidea}a).
In such a system the polariton wave function contains a component
originating from exciton in two different quantum wells and from the two coupled
optical modes. As a result, the polariton relaxation from a level with a
dominant contribution by the exciton from one quantum well to a level with
a dominant contribution by the exciton from another well is accompanied by
a change in the spatial position of the exciton. In this way, the polariton
relaxation is accompanied by energy transfer between
different, separated in space, quantum wells. {\color{black}Doping of the quantum wells in one of the
microcavities with a small amount of Mn$^{2+}$ ions enhances the exciton Zeeman splitting in magnetic field due to the
{\it s,p-d} exchange interaction between the extended band states and the
localized spins of the Mn$^{2+}$ ions (see Figure~\ref{theidea}b).\cite{Furdyna:JAP1988} To maximize exciton splittings we choose the Faraday geometry with magnetic field applied along the sample growth direction, thus parallel to the quantization axis. In consequence, we are able to tune the energy of the excitons in the Mn-doped quantum well ($QW_{\text{DMS}}$) below or above
the energy of the excitons in the non-magnetic quantum well ($QW$) using
a magnetic field.\cite{Sadowski:ThinSF1997}} Since the exciton density transfer between the wells is assisted by
energy relaxation, by adjustment of the relative energy order of the excitons in
the $QW_{\text{DMS}}$ and $QW$ one gains control of the direction of the
transfer, as schematically shown in Figure~\ref{theidea}a.

\section*{Results}
\subsection*{Four-level polaritonic system}

\begin{figure}
   \includegraphics[width=0.85\linewidth]{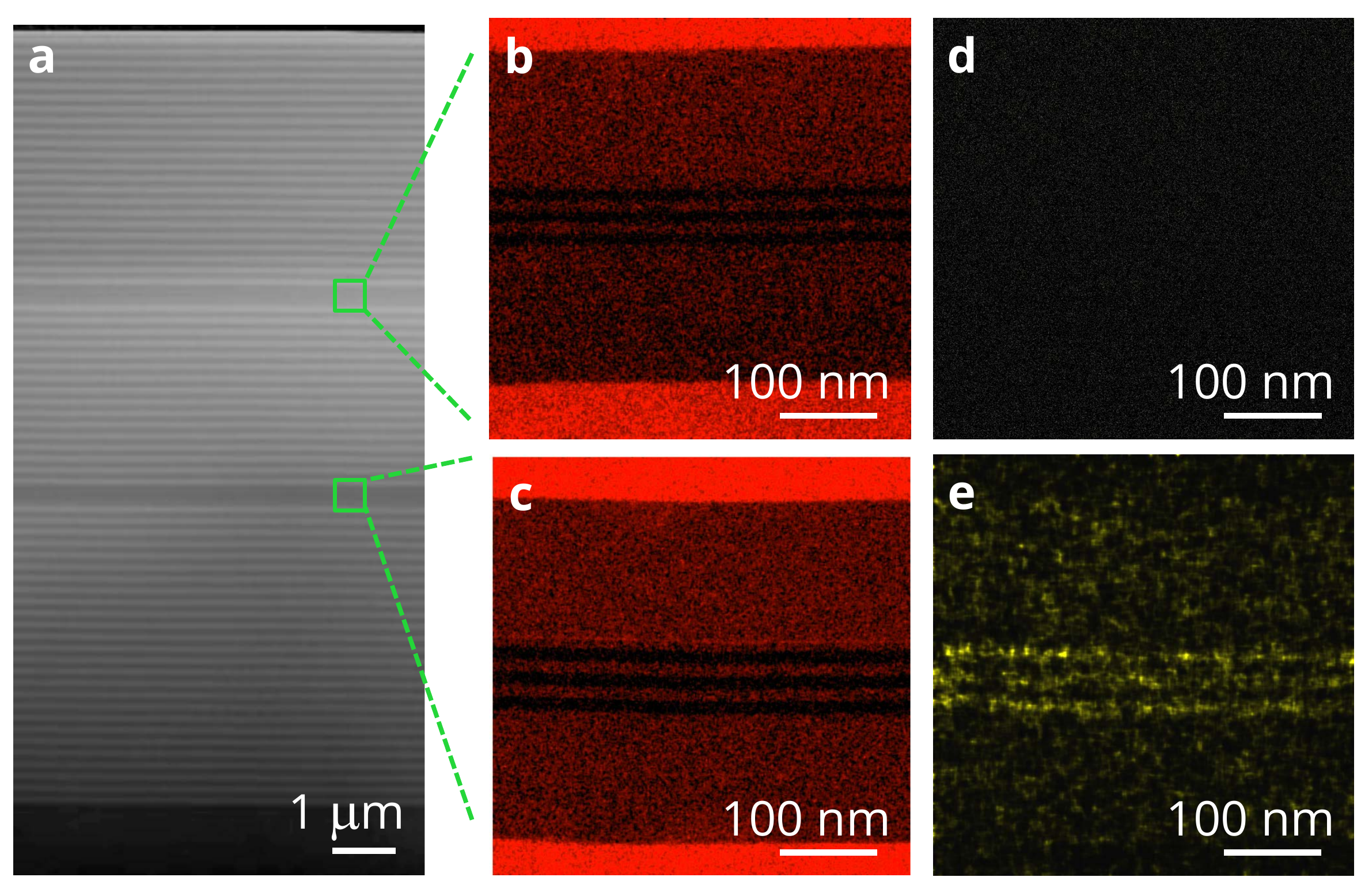}
\caption{{\bf Transmission Electron Microscopy images of {\color{black}double coupled microcavities} cross-section.} {\bf a} The structure comprises two (Cd,Zn,Mg)Te microcavities coupled by a semi-transparent Bragg mirror.
Three (Cd,Zn)Te and three (Cd,Mn,Zn)Te quantum wells are embedded in the upper
and lower microcavities, respectively, as confirmed by the
Energy-Dispersive X-ray spectroscopy investigations of spatial
distribution of Mg atoms (panels {\bf b} and {\bf c}) and Mn atoms (panels {\bf d} and {\bf e}) in the sample cross-section.}
  \label{TEM}
\end{figure}

As shown in Figure~\ref{TEM}a-c, three 10~nm wide non-magnetic (Cd,Zn)Te quantum wells and three 12~nm
wide, Mn$^{2+}$ doped (semimagnetic) (Cd,Mn,Zn)Te quantum wells are, as intended, embedded
in the centre of the upper and lower microcavity, respectively. Incorporation of the Mn$^{2+}$ ions only in the
$QW_{\text{DMS}}$ is confirmed by the Transmission Electron Microscopy images shown in Figure~\ref{TEM}d-e.

\begin{figure}
   \includegraphics[width=0.85\linewidth]{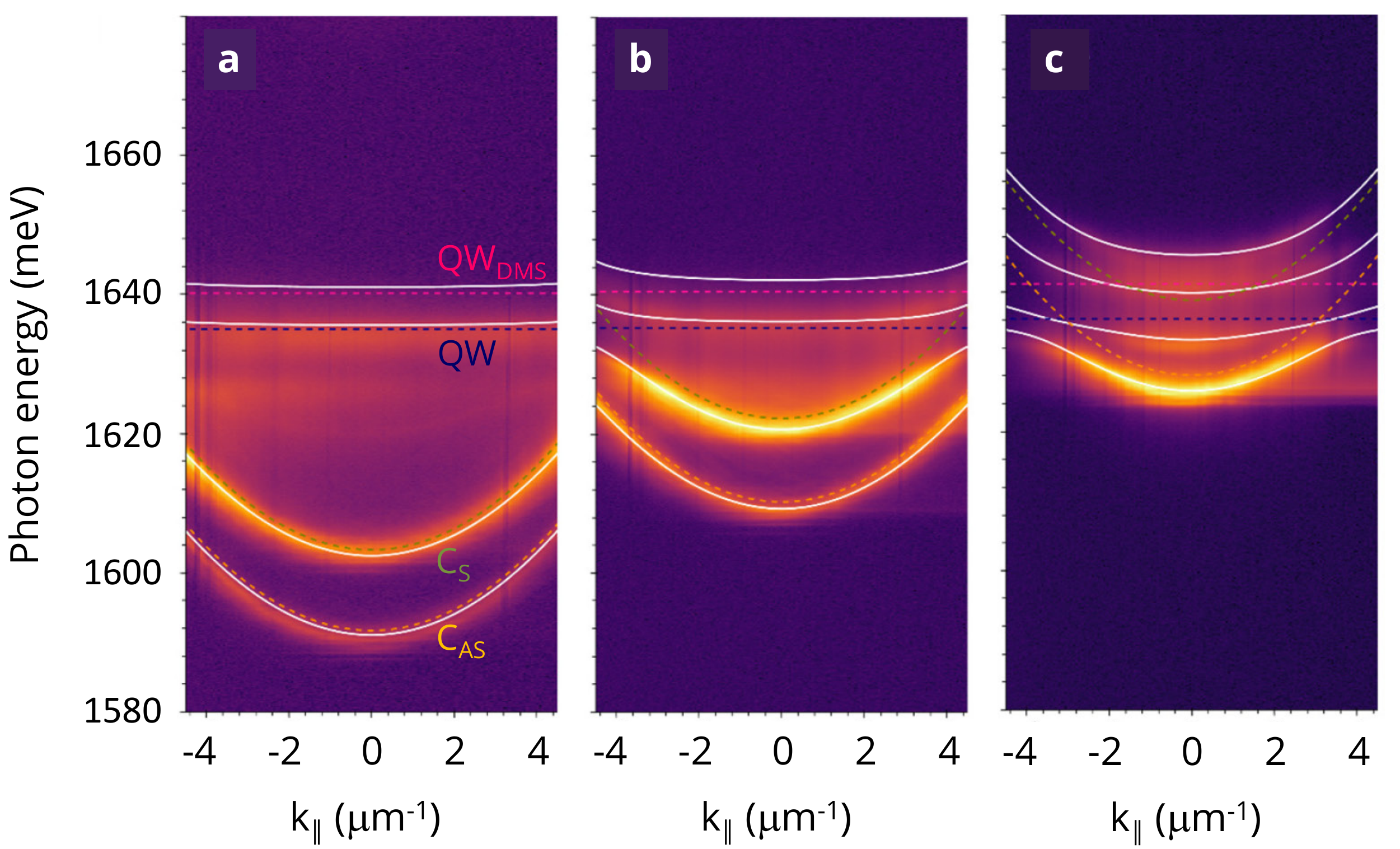}
\caption{{\bf Formation of a four-level polariton system {\color{black}in double coupled microcavities with quantum wells}.} Emission spectra
resolved in in-plane photon momentum k$_{\parallel}$ (shown on a
logarithmic intensity scale). Dotted lines represent the uncoupled exciton
states in the $QW$ and $QW_{\text{DMS}}$, as well as the optical modes
$C_{\text{S}}$ and $C_{\text{AS}}$ delocalized spatially over the two coupled
microcavities. Solid lines represent polariton levels calculated as
eigenvalues of the Hamiltonian $H$ (Eq.~\ref{Hamiltonian}). Detuning
between the optical modes and exciton levels increases from {\bf a} to {\bf
c}.}
  \label{kspace}
\end{figure}

In order to prove evidence for the formation of a four-level
polaritonic system in the studied sample, in Figure~\ref{kspace} we show
emission spectra of the structure resolved in in-plane photon momentum
space k$_{\parallel}$. The spectra are registered at consecutive
positions on the sample along the direction of the microcavity thickness
gradient, i.e.\ as a function of the detuning between the coupled, symmetric
($C_{\text{S}}$) and antisymmetric ($C_{\text{AS}}$), optical modes of the structure and the $QW$
and $QW_{\text{DMS}}$ excitons. For an even number of Bragg pairs separating
the microcavities, as in the present case, the $C_{\text{S}}$ mode is \emph{higher} in
energy than the $C_{\text{AS}}$ one.

The case of a large negative detuning between the excitons and
microcavity modes is shown in Figure~\ref{kspace}a. Exciton-photon mixing
is negligible here, and individual components contributing to the four-level
polaritonic system are clearly distinguished. The approximately parabolic
dispersion identifies the $C_{\text{S}}$ and $C_{\text{AS}}$ modes. The $QW$ transition with
a negligible dispersion is seen at around 1640~meV, while the
$QW_{\text{DMS}}$ one occurs at 1645~meV.
Excitation energy below the bandgap energy of any layer of the structure except the $QW$ and $QW_{\text{DMS}}$ ensures that the excitation penetrates both microcavities.
The absorption of the quantum wells in the upper microcavity is below 10\%, which
means that carriers are generated in the quantum wells in the upper and
lower microcavities with comparable efficiency. Emission from the $QW_{\text{DMS}}$ is
vanishingly weak, however, due to its filtering by the stopbands of the middle and
upper Bragg mirrors. When the coupled modes approach the excitonic levels,
anticrossing is observed (see Figure~\ref{kspace}b), which testifies to
strong light-matter coupling conditions. Figure~\ref{kspace}c shows the
case of modes-excitons resonance. In contrast to the case shown in
Figure~\ref{kspace}a, all four emitting levels exhibit a clear dependence
on k$_{\parallel}$, proving a non-negligible contribution by the
photonic part to each state of the four-level polaritonic system.

In order to describe the observed polariton dispersion quantitatively, we
introduce the Hamiltonian $H$ (Eq.~\ref{Hamiltonian}) representing a four
coupled oscillator
model \cite{Panzarini:PRB1999,Richter:APL2015,Dufferwiel:APL2015}. The
Hamiltonian $H$ written in the basis of the exciton states $QW$ and
$QW_{\text{DMS}}$, with mode $M$ localized in the upper and mode
$M_{\mathrm{DMS}}$ localized in the lower microcavity, takes the form:

\begin{equation}
  H=
  \tikz[baseline=(M.west)]{%
    \node[matrix of math nodes,matrix anchor=west,left delimiter=(,right
delimiter=),ampersand replacement=\&, minimum width=4em] (M) {%
    QW           \& \Omega/2 \& 0                  \& 0 \\
    \Omega/2 \& M              \& \kappa/2   \& 0 \\
    0                \& \kappa/2 \& M_\mathrm{DMS} \& \Omega_\mathrm{DMS}/2
\\
    0      \& 0      \& \Omega_\mathrm{DMS}/2 \& QW_\mathrm{DMS}\\
    };
corners=3pt,color=blue] {};
dashed] {};
color=red] {};
  }.
  \label{Hamiltonian}
\end{equation}

Off-diagonal elements of the matrix represent couplings in the system,
$\kappa$ describes the strength of the interaction between the (degenerate) $M$
and $M_{\mathrm{DMS}}$ giving rise to the emergence of the $C_{\text{S}}$ and $C_{\text{AS}}$
optical modes \cite{Stanley:APL1994,Armitage:PRB1998,
Panzarini:PRB1999,Sciesiek:CGD2017} delocalized spatially over the two
microcavities (the respective squared electric field distributions for the
$C_{\text{S}}$ and $C_{\text{AS}}$ modes are shown in Figure~\ref{theidea}a),
$\Omega$ is the coupling constant between the $QW$ exciton and the mode $M$
of the upper microcavity and $\Omega_\mathrm{DMS}$ represents the coupling
between the $QW_{\text{DMS}}$ exciton and the $M_{\mathrm{DMS}}$ mode of the lower
microcavity. Direct coupling of the excitons with the optical mode
confined in the adjacent microcavity is
neglected \cite{Panzarini:PRB1999,Richter:APL2015,Dufferwiel:APL2015}.

Fitting the energies obtained from diagonalization of the Hamiltonian $H$ to
the energies of the optical transitions reported in Figure~\ref{kspace} determines
the value of $\kappa$ to be 13~meV. Such a value is, in fact, expected for a
microcavity separation of 16~{\color{black} Distributed Bragg Reflector~(DBR)}
pairs,\cite{Bayindir:JOptics2001,Sciesiek:CGD2017} as in the present case.
The vacuum Rabi splitting $\Omega$ for the $QW$ exciton is $(10.0~\pm~0.4)$~meV, and
$\Omega_{\text{DMS}}$ for the $QW_{\text{DMS}}$ exciton is $(12.5~\pm~0.4)$~meV. In
simulations, only the bare level energies are changed by detuning, while the
coupling constants remain fixed (they change by less than 10\% when changing the
position on a 7~mm $\times$ 20~mm sample). In terms of the splitting energy per quantum
well, the values obtained are consistent with previous reports on II-VI
polariton systems \cite{Kasprzak:Nature2006, Sebald:APL2012, Klein:APL2015,
Rousset:APL2015, Sawicki:CommPhys2019}.

\begin{figure*}[h!]
\includegraphics[width=0.9\linewidth]{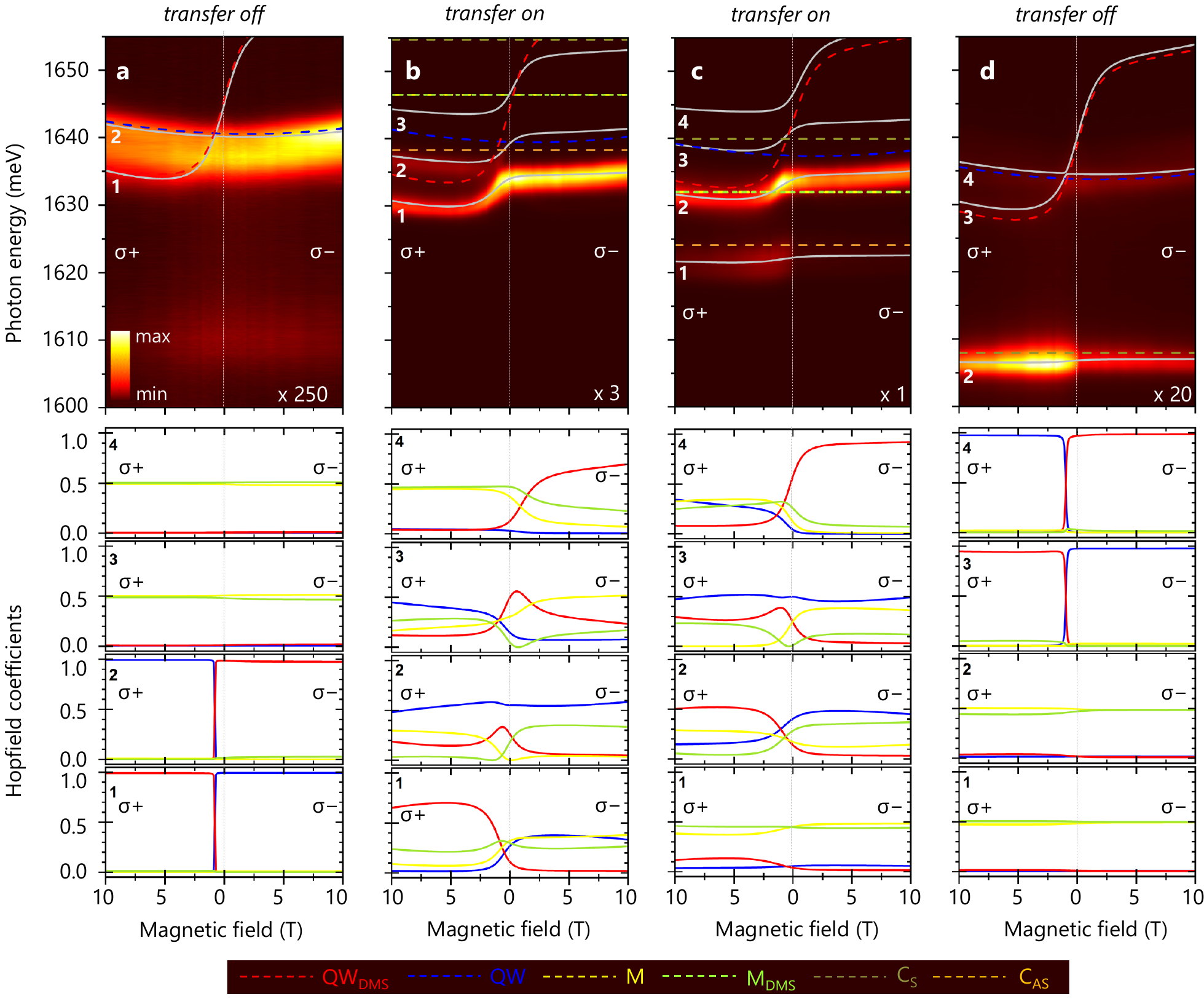}
\caption{{\bf Magnetic field controlled, polariton-mediated energy transfer between macroscopically distant quantum wells.} Photoluminescence spectra as a function of the magnetic field for consecutive values of the detuning
between the microcavity modes and quantum well excitons, decreasing from {\bf
a} to {\bf d}. Circular polarizations of the detection are indicated. The bare levels of the excitons in the $QW$ and QW$_{\text{DMS}}$ (simulated following Eq.~\ref{QW:energy} and \ref{QWDMS:energy}, respectively), as well as the optical modes M and M$_{\text{DMS}}$ of the upper and lower microcavities are shown by the dashed blue, red, yellow, and light-green lines, respectively. Delocalized optical modes $C_{\text{AS}}$ and $C_{\text{S}}$ arising from the
coupling between $M$ and $M_{\text{DMS}}$ are denoted by the dashed orange
and olive lines, respectively. Calculated polariton levels are shown
by solid white lines. Hopfield coefficients \emph{vs} magnetic field
determined in the basis of the $QW_{\text{DMS}}$, $QW$, $M$ and
$M_{\text{DMS}}$ for polariton levels from 1 to 4 are provided below the
photoluminescence maps with the respective colour code.
The \emph{"transfer on"} case, where the coupled modes are tuned to resonance with the $QW$ and $QW_{\text{DMS}}$ excitons enabling long-distance, polariton-mediated energy transfer is shown in panels {\bf b} and {\bf c}. The \emph{"transfer off"} case, where the coupled modes are detuned with respect to the excitons, which precludes the interaction and energy transfer between magnetic and non-magnetic exciton states is shown in panels {\bf a} and {\bf d}.
}
\label{PLvsB}
\end{figure*}

We note that the coupling constants are of the same order as the energy
separation between the $QW$ and $QW_{\text{DMS}}$ excitons. This means that the
coupling with the delocalized optical modes of the structure indeed ensures
hybridization of the $QW$ and $QW_{\text{DMS}}$ states, being a prerequisite
for efficient polariton-mediated energy transfer between the quantum wells.

{\color{black}In previous studies, the magnetic field applied in Faraday geometry enabled lowering of the polariton condensation threshold\cite{Rousset:PRB2017} or, when applied in the Voigt geometry, controlling of the polariton condensate propagation in single microcavities.\cite{Caputo:CommunPhys:2019}
Here, we demonstrate photon-mediated exciton interaction and energy transfer between the macroscopically distant quantum wells in double microcavity structure and use the magnetic field as a mean to control of the transfer direction.}

\subsection*{Photon-mediated interaction between magnetic and non-magnetic exciton}
Figure~\ref{PLvsB} shows the emission spectra resolved in circular
polarization, registered as a function of magnetic field for consecutive detunings between the microcavity
modes and quantum well excitons decreasing from panel a to d. Corresponding
Hopfield coefficients \cite{Hopfield:PR1958} determined from the
diagonalization of the Hamiltonian $H$ (Eq.~\ref{Hamiltonian}) are shown
below the spectra (note the labels run from 1 to 4 in order of increasing level energy).
Below we analyze the emission energy and intensity of the
dependencies observed in the experiment.

The non-resonant case, where the coupled modes are far detuned to higher energy with respect to the quantum well excitons (around 20~meV at $B =
0$~T), is shown in Figure~\ref{PLvsB}a. In this case, the observed emission arises predominantly from the exciton confined in the $QW$, as indicated by only a slight change in the emission energy and intensity upon application of the
magnetic field (at $B = 10$~T $\sigma-$ intensity $I^{\sigma-}$ is higher than the $\sigma+$ intensity $I^{\sigma+}$ by a factor of $\lesssim$2. Such a small splitting, as well as only weak (below 30\%) and negative polarization of the emission (=($I^{\sigma+}-I^{\sigma-})/(I^{\sigma+}+I^{\sigma-}))$ is expected in the case of a non-magnetic quantum well, since only linear Zeeman splitting (excitonic g-factor of 0.7, consistent with
Ref.~\citenum{Zhao:APL1996}) and diamagnetic shift ($\gamma$ = 0.018), both
of relatively small magnitude, affect the exciton energy $E_{QW}^{\sigma\pm}(B)$:
\begin{align}
E_{QW}^{\sigma\pm}(B) = E_{QW}(B = 0) \pm g\mu_{B} B + \gamma B^2
\label{QW:energy}
\end{align}
In the case of the $QW_{\text{DMS}}$, the exciton energy $E_{QW_{\text{DMS}}}^{\sigma\pm}(B)$ varies much more upon application of magnetic field due to an additional contribution coming from the {\it s,p-d} exchange interaction between the exciton spin and the localized 5/2 spins of the Mn$^{2+}$ ions:
\begin{align}
E_{QW_{\text{DMS}}}^{\sigma\pm}(B)= E_{QW_{DMS}}(B = 0) \mp \frac{E_{sat}}{2} B_{5/2}\left(
\frac{\frac{5}{2} g_{\mathrm{Mn}} \mu_\mathrm{B} B}{k_\mathrm{B} T_{\mathrm{eff}}} \right) \pm g\mu_{B} B + \gamma B^2.
\label{QWDMS:energy}
\end{align}
In Eq.~\ref{QWDMS:energy} the term $E_{sat} = x_{\mathrm{Mn}} (N_0 \alpha - N_0 \beta) S_{0}$ represents exciton giant Zeeman splitting at saturation, defined by the Mn dopant concentration $x_{\mathrm{Mn}}$, the \emph{s - d} and \emph{p - d} exchange integrals $N_0 \alpha = 0.22$~eV and $N_0 \beta = -0.88$~eV, respectively for electrons and holes, and $S_{0} = 2.12$ being the effective spin of the Mn$^{2+}$ ion in (Cd,Mn)Te.\cite{Gaj:SSC1979} The $B_{5/2}$ is the modified Brillouin function\cite{Gaj:SSC1979} with parameters: Mn$^{2+}$ ion Land\'{e} factor $g_{\mathrm{Mn}}$ = 2, Bohr magneton $\mu_\mathrm{B}$, effective temperature of the Mn spins $T_{\mathrm{eff}}$ = 2.3~K, and the Boltzmann constant $k_\mathrm{B}$.

It is worth to note, that crossing of the $QW$ and $QW_{\text{DMS}}$ exciton levels occurring at around $B_{crit}$
= 0.8~T at $\sigma+$ polarization in Figure~\ref{PLvsB}a does not affect the energy of the emission. This indicates that when the mode is far detuned from the excitons, no
interaction between excitons in the $QW$ and $QW_{\text{DMS}}$ occurs. The
negligible hybridization of the $QW$ and $QW_{\text{DMS}}$ states in the
non-resonant case is reflected by a sharp exchange of the $QW$ and
$QW_{\text{DMS}}$ components in the polariton states 1 and 2 when the magnetic
field passes $B_{crit}$ (see the Hopfield coefficients for levels 1 and 2 in
Figure~\ref{PLvsB}a). The emission from the $QW_{\text{DMS}}$ in the non-resonant case is filtered out by the middle and upper DBRs. Thus despite being excited by the laser, it is not manifested in the spectrum. In the non-resonant case, the emission spectral width is just the linewidth of the $QW$ confined neutral exciton transition. It is larger than the radiative lifetime defined limit due to a sample structural disorder. An additional broadening of the emission towards lower energy seen in Figure~\ref{PLvsB}a is attributed to a presence of transition of charged exciton\cite{Kossacki:JPCM2003,Muszynski:APL2020} confined in the QW.

The resonant case, where either the $C_{\text{AS}}$ or $C_{\text{S}}$ mode is tuned to the
close spectral vicinity of the quantum well levels, is shown in
Figure~\ref{PLvsB}b or \ref{PLvsB}c, respectively. The spectra reveal an
anticrossing of the polariton states for which a contribution from the $QW$
and $QW_{\text{DMS}}$ exciton dominates (they are, respectively, levels 1
and 2 in Figure~\ref{PLvsB}b and levels 2 and 3 in
Figure~\ref{PLvsB}c). The anticrossing occurs at polarization $\sigma+$ at
$B_{crit}$ = 0.8~T, where the uncoupled $QW$ and $QW_{\text{DMS}}$
exciton levels cross, proving the photon-mediated interaction between the macroscopically distant, magnetic and non-magnetic, excitons. The Hopfield coefficients confirm enhanced hybridization of the $QW$ and
$QW_{\text{DMS}}$ excitons in the resonant case. In the resonant case, the  emission spectral width is the average of the $QW$ exciton and the mode linewidths weighted by the respective contributions of the two components to the emitting state.
In the following discussion by conditions "below $B_{crit}$" we mean $B > B_{crit}$ in $\sigma+$ polarization, while by "above $B_{crit}$" we mean the remaining field range and polarization of the signal (i.e., $B < B_{crit}$ in $\sigma+$ and $B > 0$~T in $\sigma-$ polarization).

\subsection*{Magnetic field controlled, polariton-mediated energy transfer over a macroscopic distance}

Let us now consider the emission intensity dependences, starting from the
resonant (\emph{"transfer on"}) case shown in Figure~\ref{PLvsB}b. Below $B_{crit}$, the dominant
contribution to level 1 comes from the exciton confined in the
$QW_{\text{DMS}}$, while the dominant contribution to level 2 comes
from the exciton confined in the $QW$. Quantum wells in both microcavities
are excited with comparable intensity, and the $QW_{\text{DMS}}$ emission
is not as efficiently filtered out by the DBRs stopband as it was in the
non-resonant case. Thus, if there were no exciton density transfer between the $QW$ and
$QW_{\text{DMS}}$ one would expect a comparable emission intensity from
levels 1 and 2. This is clearly not the case: the emission occurs
mostly from the lowest in energy, polariton level 1. The observed enhancement
of the population of level 1 and depletion of the population of
level 2 reflects the efficient relaxation of the polaritons to level 1 from
level 2. This points towards efficient exciton density transfer from the $QW$ placed in the upper microcavity to the $QW_{\text{DMS}}$ placed in the lower microcavity, 2.15~$\mu$m away.
The observed behaviour, in particular the depletion of the polariton level
2, cannot result from intrawell spin relaxation of the exciton in the $QW$,
since its spin polarization is weak (see Figure~\ref{PLvsB}a) and it
is not affected by a superposition with the optical mode (at least below
the polariton condensation density, as in the present
case).\cite{Fischer:PRL2014,Sturm:PRB2015,Krol:PRB2019}

Exciton density transfer between the $QW$ and $QW_{\text{DMS}}$ does not require
exciton spin relaxation. It is assisted, however, by its energy relaxation.
The energy difference between the $QW$ and $QW_{\text{DMS}}$ excitons is
relatively small (up to 5~meV, depending on the magnetic field), which
facilitates its efficient dissipation by acoustic phonons and ensures high
exciton density transfer rates \cite{Price:AnnPhys1981}. Measurements of time-resolved
micro-photoluminescence from the cleaved edge of the sample indicate that
the lifetime of the exciton in either the $QW$ or $QW_{\text{DMS}}$ is around
200~ps (not shown). We thus state that the transfer time is much
smaller than this value, in the range of tens of ps or shorter.

In turn, above $B_{crit}$ in Figure~\ref{PLvsB}b the contribution to the lowest polariton level 1 from the $QW$ exciton dominates over the contribution from the $QW_{\text{DMS}}$ exciton.
A dominant population of level 1 above $B_{crit}$ indicates that
the direction of the transfer is reversed and the exciton density shifts
from the $QW_{\text{DMS}}$ to the $QW$. Here, due to the large splitting of the
$QW_{\text{DMS}}$ exciton for $B > 0$~T, an additional contribution to the
depletion of the $\sigma-$ polarized exciton in the $QW_{\text{DMS}}$ exciton
appears, resulting from spin and energy exciton relaxation within the
$QW_{\text{DMS}}$\cite{Furdyna:JAP1988}. The excitation, kept constant during the magnetic field dependent measurements, is linearly polarized and non-resonant, so that it does not introduce any imbalance between $\sigma+$ and $\sigma-$ populations of the photocreated excitons.

The energy and intensity dependences of the resonant (\emph{"transfer on"}) case
presented in Figure~\ref{PLvsB}c are similar to the case in
Figure~\ref{PLvsB}b. Excitons contributing to polaritons on the most
populated level 2 originate predominantly either from the $QW$ (for $B$ below
$B_{crit}$) or the $QW_{\text{DMS}}$ (for $B$ above $B_{crit}$).
Comparable dependences in Figure~\ref{PLvsB}b and Figure~\ref{PLvsB}c point
towards their similar interpretation and indicate that both coupled modes
$C_{\text{S}}$ and $C_{\text{AS}}$ mediate the energy transfer between the quantum wells
with comparable efficiency. Such a result is consistent with comparable
intensities of the electric field associated with $C_{\text{S}}$ and $C_{\text{AS}}$ at the
centres of the microcavities, as seen in Figure~\ref{theidea}a. Analysis of
the Hopfield coefficients in Figure~\ref{PLvsB}c reveals that feeding of
the polariton levels indeed occurs through the quantum well excitons, as
polariton level 2, of strongly excitonic character, is much more populated
than the lowest, but mostly photonic in nature, level 1. Consistently, emission from the level 1 in $\sigma+$ polarization, where a net exciton component is present (see the Hopfield coefficients) is stronger than in $\sigma-$ polarization, where the exciton content to level 1 is negligible.

Finally, Figure~\ref{PLvsB}d shows that when the optical modes are far detuned
to lower energy with respect to the quantum well excitons (\emph{" transfer} off" case), the ordinary
crossing of exciton energy levels of quantum wells occurs. This indicates
that the photon-mediated interaction and energy transfer between the $QW$ and $QW_{\text{DMS}}$ exciton states have been turned off again. Also in this case, a net exciton content results in enhancement of the $\sigma+$ polarized with respect to $\sigma-$ polarized emission of the level 2.

\section*{Discussion}
Photon-mediated hybridization of excitonic states and enhanced energy transfer
efficiency over distances of the order of 100~nm have been studied so far
in microcavities with embedded layers of organic
molecules \cite{Lidzey:Science2000,Lodden:PRL2012,
Coles:NatureMat2014,Orgiu:NatureMat2015,Feist:PRL2015,Schachenmayer:PRL2015,Zhong:AngewandteChem2016,Zhong:Angewandte2017} or hybrid structures involving perovskite or two-dimensional transition metal
dichalcogenides layers \cite{Wenus:PRB2006,Lanty:PRB2011,
Slootsky:PRL2014,Flatten:NatureComm2017,Waldherr:NatComm2018}. Very recently,
use of a hybrid organic-inorganic coupled microcavity enabled the energy transfer on a distance reaching 1.5~$\mu$m \cite{Jayaprakash:LSA2019}.

A long-distance, spin-dependent interaction and polariton-mediated energy transfer between quantum emitters demonstrated in the present work are
promising for deterministic designing of novel
photonic devices such as, e.g., bosonic Josephson
junctions \cite{Albiez:PRL2005, Lagoudakis:PRL2010,Abbarchi:NaturePhys2013},
so far limited to disordered polaritonic systems.
Formation of spatially delocalized polaritons in a coupled multi-cavity structure and the possibility of their
manipulation by a magnetic field makes the presented system useful for
studies and implementation in the recently intensively developing areas of
quantum polariton networks \cite{Liew:PRB2018}, quantum
simulators \cite{Berloff:NatureMat2017} and production of hyper-entangled
photons with a strongly reduced noise background \cite{Portolan:NJofPhys2014}.

The present study has been performed in a linear regime, where interactions
between the polaritons are negligible due to their small density. An exciting extension of this work would be a study of the presented
four-level polaritonic system to the high excitation limit, where stimulated
scattering in a non-linear regime induces such effects as the polariton
Bose-Einstein condensation. Such a study should not only answer the
question whether the condensate can boost a long-distance energy transfer in a semiconductor, but it should also enable observation of new phenomena, such as magnetic
field tunable polariton parametric scattering between polariton branches or
switchable, multiple wavelength polariton lasing. Enhancement of optical
nonlinearities thanks to the exciton hybridization should enable an
ultra-low threshold for lasing.

\section*{Conclusions}
We have accomplished magnetic field controlled coupling and energy transfer between semiconductor quantum wells over a distance exceeding 2~$\mu$m. The distant interaction is ensured by the strong coupling of the excitons to a common optical mode
delocalized over the two coupled microcavities. The magnetic field enables
control of the strength of the coupling and the energy order of the semimagnetic
quantum well exciton with respect to the non-magnetic quantum well exciton.
This makes possible control of the direction of the polariton-mediated energy transfer by
an external magnetic field, which is promising for magnetic field
controlled routing of the excitation in a semiconductor. The multilevel system presented is attractive for further studies of nonlinear polariton effects and for a
wide range of applications in optoelectronics and quantum information
science.

\section*{Methods}

\subsection*{Sample}
The structures are grown by molecular beam epitaxy on a ZnTe buffer layer
deposited on a (100) oriented GaAs substrate, with real-time monitoring
of the thickness of the layers by \emph{in situ} reflectivity during around
12 hours of growth. The structure contains two (Cd,Zn,Mg)Te 3$\lambda$/2
microcavities ($n_{cav}$ = 2.7 at $\lambda$ = 750 nm). The microcavities
are sandwiched between and separated by DBRs
made of $30$, $16$ and $28.5$ pairs of alternating refractive index layers
in the case of the lower, the middle and the upper mirrors, respectively (see
Figure~\ref{TEM}a). High ($n_{high} = 2.95$ at 750~nm) and low ($n_{low} =
2.4$ at 750~nm) refractive index layers are made of (Cd,Zn,Mg)Te with,
respectively, 10\% and 50\% Mg content. Both microcavities are of wedge type,
which results in a gradient of the mode energy of $10.5$ meV/mm, as deduced
from reflectivity spatial mapping. At every point on the sample the
microcavities share the same thickness, which ensures a maximum degree of
coupling between their optical modes. The quality factor of the microcavities deduced
from the coupled modes linewidth observed in reflectivity exceeds 2000.
The energy gradient of a quantum well confined exciton amounts to around $0.6$
meV/mm.
The concentration of the manganese in $QW_{\text{DMS}}$ is determined to 0.8\% by fitting a modified Brillouin
function\cite{Gaj:SSC1979} to the $QW_{\text{DMS}}$ exciton shift in
the magnetic field.

\subsection*{Experiment}
Photoluminescence is non-resonantly \emph{cw} excited at 1.81~eV
($\lambda_{exc} = 685$~nm).
For far-field distribution of the emission measurements, the sample is
placed inside a helium-flow cryostat at 8~K. The laser beam is focused to a
1-2 $\mu$m diameter spot at the sample surface using a microscope objective
({\color{black} numerical aperture of} 0.7). The in-plane photon momentum k$_{\parallel}$ is registered in
the range from -4.5~$\mu$m$^{-1}$ to 4.5~$\mu$m$^{-1}$ by imaging the
Fourier plane of the microscope objective on the entrance slit of the
spectrometer.
For magnetic field dependent measurements, the sample is placed inside a
pumped helium cryostat at 1.8~K equipped with a superconducting split-coil
magnet producing a magnetic field of up to 10~T. The measurements are
performed in the Faraday configuration. The combination of a quarter-wave
plate and a linear polarizer allows for the detection of two circular
polarizations. The laser beam is focused to a 0.1~mm diameter spot at the
sample surface using a 500~mm focal length lens. The lens is
mounted on an automated XY translation stage, which enables spatial
mapping of the {\color{black} photoluminescence} with a step of 0.01~mm. The range of registered
k$_{\parallel}$ vectors is limited to 0.42~$\mu$m$^{-1}$ in this case. A
CCD camera coupled to a grating spectrometer serves as a detector (0.1~meV
of the overall spectral setup resolution).

Transmission Electrom Microscopy (TEM) observations are conducted on a FEI Talos F200X microscope operating at 200~kV.  The measurements are performed in Scanning TEM (STEM) mode using a high-angle annular dark-field detector and Energy-dispersive X-ray spectroscopy on a Brucker BD4 spectrometer.
Cross-sectional TEM specimens are prepared by a standard method of mechanical pre-thinning followed by Ar ion milling.

\subsection*{TMM Simulations}
The spatial distribution of the electric field in the structure shown in
Figure~\ref{theidea}a is calculated using the transfer matrix method.
Complex refractive indices of the layers are assumed following the
literature \cite{Andre:JAP1997}.

{\color{black} \section*{Data availability}
The data that support the findings of this study are available from the
corresponding author upon reasonable request.}

\section*{Competing interests}
The authors declare no competing interests.

\section*{Acknowledgments}
We gratefully acknowledge sample preparation for the TEM measurements by
Jolanta~Borysiuk (deceased), and we dedicate this work to her. This work was
partially supported by the Polish National Science Centre under projects
DEC-2013/10/E/ST3/00215 and DEC-2017/25/N/ST3/00465. The research was carried out with the use of CePT, CeZaMat and NLTK infrastructures financed by the European Union - the European Regional Development Fund within the Operational Programme ``Innovative economy" for 2007-2013.

\section*{Author contributions statement}
M.\'{S} was involved in all steps of this work. J.S. was involved in all steps of this work apart from the sample growth. W.P. epitaxially grew the samples. K.S. and T.K participated in optical measurements. A.G. participated in design of the studies and interpretation of the results. K.So. performed TEM measurements. J.S wrote and all authors reviewed the manuscript.

\end{document}